\documentclass[a4paper, notoc]{JHEP3}
\usepackage[dvips]{graphicx}
\usepackage{amsfonts}
\usepackage{amssymb}
\usepackage{epsfig}
\usepackage{cite}

\newcommand{\be}{\begin{equation}}
\newcommand{\ee}{\end{equation}}
\newcommand{\bea}{\begin{eqnarray}}
\newcommand{\eea}{\end{eqnarray}}
\newcommand{\nn}{\nonumber}
\newcommand{\mbb}{\mathbb}
\newcommand{\ti}{\times}
\newcommand{\half}{\frac{1}{2}}
\newcommand{\mc}{\mathcal}

\newcommand{\beqa}{\begin{eqnarray}}
\newcommand{\eeqa}{\end{eqnarray}}

\newcommand{\thba}[3]{\vartheta \Big[ \! \begin{array}{c}{\phantom{}\vspace{-0.5mm} #1}%
                        \\[-0.5mm]{ #2}\end{array}  \Big] \Big( { #3} \Big)}

 \title{Brane-Antibrane Backreaction in Axion Monodromy Inflation}
\author{Joseph P. Conlon
 \\ Rudolf Peierls Center for Theoretical Physics, 1 Keble Road \\
 Oxford OX1 3NP, UK \\ Email: \email{j.conlon1@physics.ox.ac.uk}}

\abstract{We calculate the interaction potential between D5 and $\overline{\hbox{D}}5$ branes
wrapping distant but homologous 2-cycles. The interaction potential is logarithmic in the separation
radius and does not decouple at infinity. We show that logarithmic backreaction is generic for 5-branes wrapping distant but
 homologous 2-cycles, and we argue that this destabilises models of axion monodromy inflation involving
NS5 brane-antibrane pairs in separate warped throats towards an uncontrolled region.}

\preprint{}

\begin{document}

\section{Introduction and Review}

Inflation is a powerful and compelling explanation for both the large-scale homogeneity and flatness of the universe and also for the origin
of the small density perturbations that lead to the growth of structure.
Among physical processes for which there is at least some degree
of observational evidence, inflation also probes the largest energy scales. This makes it natural to attempt to build models
of inflation within string theory, the leading candidate theory of Planck scale physics. Such model building can suggest both
novel signatures (for example relic cosmic superstrings from brane/antibrane annihilation) and 
constraints on inflationary parameter space.
Reviews of inflationary model building in string theory include \cite{09010265, 11082659, 11082660}.

One potential constraint that has emerged is the possibility that string theory forbids models with large tensors.
This is equivalent to the statement that string theory does not admit inflationary potentials which are flat over trans-Planckian
field ranges. It is empirically true that the majority of string inflation models involve small field inflation, and many fields
in string models cannot be moved through a Planckian distance without going to a regime where control is lost.
For example, for internal moduli a Planckian displacement drives the Calabi-Yau close to a degenerate limit.

There are nonetheless some interesting proposals for string inflation models with observable tensors, for example
 \cite{08033085, 08080706} or \cite{08080691}. However inflationary models require the highest levels of complexity in string model building -
 even prior to considering flatness of the potential, they require as a prerequisite
 moduli stabilisation, supersymmetry breaking and approximately de Sitter vacua.
They are also (by necessity) far from the regions of moduli space where direct and precise computations are easily carried out,
for example through worldsheet techniques. It is therefore never easy to demonstrate full consistency of any proposed model
of string inflation.

Given the potential observational significance of large tensors it is important to perform a 
close examination of proposed models with large tensors. 
In this paper we will study a candidate model of large field inflation, axion monodromy inflation \cite{08033085, 08080706, 09072916, 09121341, 11103327}. In particular we claim that the effects of brane backreaction in these models are more serious than previously estimated.

Let us first give a brief review of the models of axion monodromy inflation (for full details see \cite{08080706, 09072916}).
These models are formulated in the context of type IIB
flux compactifications with nonperturbative stabilisation of the K\"ahler moduli. The required minimal geometry
and brane configuration is shown in figure 1.
\begin{figure}
\begin{center}
\epsfig{file=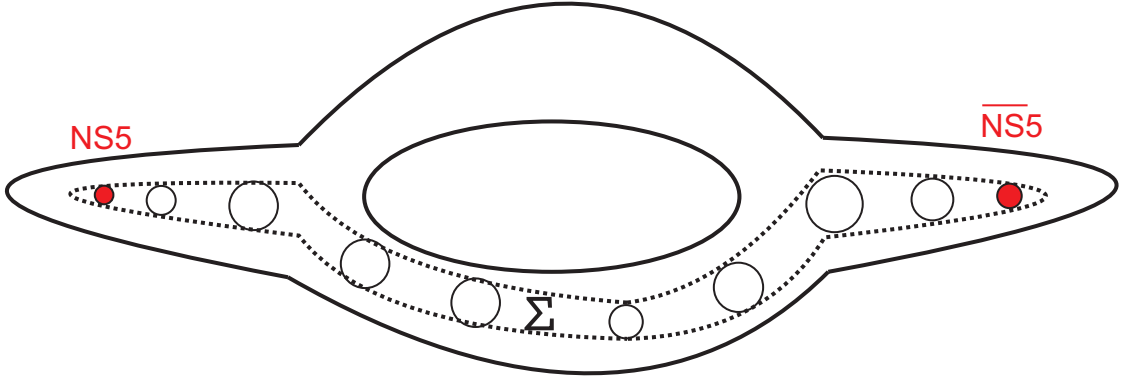,height=5cm}
\caption{The schematic brane configuration required for models of axion monodromy inflation. There is one NS5 brane and
one $\overline{\hbox{NS}}5$ brane down separate warped throats, wrapping the homologous 2-cycle $\Sigma$. Inflation is driven by the vev
$c \equiv \int_{\Sigma} C_2$ of a RR 2-form on this cycle.}
\label{z4orbifold2h}\end{center}\end{figure}
 It consists of an NS5 brane down a warped throat, wrapping a 2 cycle $\Sigma$. In a distant throat - but containing a homologous 2 cycle - there is an anti-NS5 brane, which wraps the homologous cycle and ensures tadpole cancellation.
Tadpole cancellation for this 2-cycle requires the presence of both the NS5 and the $\overline{\hbox{NS}}5$
brane.\footnote{The use of NS5 branes rather than D5 branes arises from the detailed form of nonperturbative corrections to the superpotential,
but this distinction will not play a material role in our analysis.}

Inflation comes from winding up the vev of $c = \int_{\Sigma} C_2$ on this 2-cycle. Through the NS5 equivalent of the DBI action,
$\int d^4 x \, e^{-\phi} \sqrt{g + g_s C_2}$, winding up $c$ increases the energy associated to the NS5 brane. The inflaton is precisely
the mode $c = \int_{\Sigma} C_2$. If this mode is `wound up' to large values, then for large $c$ the potential 
is linear in $c$ as $\int d^4 x \sqrt{g + g_s C_2} \to \sqrt{ l^4 + g_s^2 c^2}$. The DBI potential can then
 give large field inflation as the mode relaxes to zero vev. Corrections to the potential for $c$ are protected
 by the axion shift symmetry $\int C_2 \to
\int C_2 + 2 \pi$, which is broken only by the presence of the NS5 brane. The warped throats are necessary to ensure
the energy density of the branes is suppressed below the string scale.

Many aspects of this model are considered in \cite{08080706, 09072916}. The part we will focus on is the backreaction due to having
a NS5 and anti-NS5 pair that wrap distant but homologous cycles.
Effects of brane backreaction on the metric have been considered in \cite{08080706, 09072916}. However the analysis
there concerns the backreaction coming from the D3-brane charge and tension that is induced on the 5-brane 
from the Chern-Simons like term when $c$ is wound up.
The direct backreaction due to the 5-branes was not considered.
The implicit and apparently reasonable
assumption appears to be that a single 5-brane wrapped on a small cycle is string scale in size and
will at large transverse distances from the cycle
effectively behave as a 3-brane in terms of backreaction.

The argument we will make below is that it is the backreaction from the 5-branes that is actually most dangerous.
We claim that rather than morally being a D3-$\overline{\hbox{D}}$3 pair, the NS5-$\overline{\hbox{NS}}$5 brane pair are
morally a D7-$\overline{\hbox{D}}$7 pair: the interaction potential between
homologous NS5 and $\overline{\hbox{NS}}5$ branes is never small, grows logarithmically with the separation distance,
and does not decouple at infinity.

The structure of the paper is as follows. We first review the worldsheet calculation of the $r^{-4}$ D3-$\overline{\hbox{D}}$3 interaction potential
in flat space. We then consider orbifold models where we can perform a worldsheet calculation of the interaction potential between 
D5 and $\overline{\hbox{D}}5$ branes wrapping distant but homologous 2-cycles. We show that this interaction potential is logarithmic in the separation radius.
We explain why this result should hold more generally beyond orbifold models, and apply it to models of axion monodromy inflation.

\section{Brane-Antibrane Interaction Potentials}

\subsubsection*{D3/$\overline{\hbox{D}}$3 interaction potential in flat space}

We start with a brief review of the derivation of the D3/$\overline{\hbox{D}}$3 interaction potential in toroidally compactified flat space.
This is determined by computing the 1-loop vacuum energy in the D3/$\overline{\hbox{D}}$3 background, and can be extracted from the
 annulus partition function. This partition function is
\be
\label{pf}
\int \frac{dt}{2t} \underbrace{\frac{1}{(2 \pi^2 t)^2}}_{4-momenta}
\underbrace{\hbox{Tr}_{CP}(1)}_{Chan-Paton \, \, trace} \sum_{\alpha, \beta = 0,1/2} \eta_{\alpha \beta}
\underbrace{\left( \frac{ \thba{\alpha}{\beta}{it} }{\eta(it)^{3}} \right)^4}_{oscillators}
\underbrace{\sum_{i=1}^6 \sum_{m_i} e^{-2 \pi t (Y_i + m_i R_i)^2}}_{winding \, \, \, modes}.
\ee
Here $Y_i$ is the separation distance between the brane and antibrane and $R_i$ is the circumference of the $i$th
dimension (for convenience we assume a toroidal compactification). $Y_i$ and $R_i$ are both measured in units
of the string length, $l_s = (2 \pi \sqrt{\alpha'})$. $t$ is the modular parameter of the annulus.
We will assume that the branes are well separated, with $R > Y \gg 1$.

The sum over $\alpha, \beta$ gives the sum over spin structures and the GSO projection, with
$\eta_{\alpha \beta}$ determining the relative weightings of the different spin stuctures. For the D3/D3 system,
$\eta_{\alpha \beta} = (-1)^{2(\alpha + \beta)}$. For a D3/$\overline{\hbox{D}}$3 system the RR charge is opposite and $\eta_{\alpha \beta}
 = (-1)^{2 \alpha}$.

The theta functions are defined by
\be
\thba{\alpha}{\beta}{it} = \sum_{n=-\infty}^{\infty} e^{2 \pi i \beta (n + \alpha)} q^{- \half (n + \alpha)^2}, \qquad \hbox{ with } q = e^{- 2 \pi t},
\ee
with modular transformations
\bea
\label{mod}
\thba{\alpha}{\beta}{it} & = & e^{2 \pi i \alpha \beta} \sqrt{\frac{1}{t}} \, \thba{-\beta}{\alpha}{\frac{i}{t}}, \\
\eta(it) & = & \sqrt{ \frac{1}{t}} \, \eta \left( \frac{i}{t} \right).
\eea
The Poisson resummation of the winding modes is given by
\be
\sum_n e^{-2 \pi t (Y + n R)^2} = \frac{1}{R \sqrt{2 t}} \sum_n e^{2 \pi i n Y/ R} e^{- \frac{\pi m^2}{2 R^2 t}}.
\ee
As $R, \,  Y \gg 1$ it is clear from the partition function (\ref{pf}) that the
amplitude is exponentially suppressed for $t \gtrsim (2 \pi Y^2)^{-1}$ and we can perform a modular transformation of the
oscillator sum to the small
$t$ regime.

Using the transformation (\ref{mod}) we obtain the small $t$ expansion of the oscillator series as
\be
\hbox{D}3/\overline{\hbox{D}}3 : 32 t^4 \left( 1 + \ldots \right),
\ee
with subleading terms of order $e^{-2 \pi/t}$.

For convenience we now assume that the brane/antibrane pair is separated only in the $5$ direction, so that
$Y_5 \neq 0$ and all other $Y_i = 0$. We shall also assume that
all toroidal radii are identical, $R_i = R, \, \, i=1, \ldots, 6$. None of these assumptions 
materially affect the physics of the model.
In the $t \ll 1$ regime the partition function is then
\bea
Z_{3\bar{3}} & = & \int \frac{dt}{2t} \frac{1}{(2 \pi^2 t)^2} \hbox{Tr}(1) \ti 32 t^4 \ti \left[ \sum_{i=1,2,3,4,6} \sum_{n_i} e^{-2 \pi t n_i^2 R^2}
\sum_m e^{-2 \pi t (Y_5 + m R)^2} \right] \nn \\
& = &
\frac{32}{2 (2 \pi)^2} \int dt \, t \, \hbox{Tr}(1) \left[ \frac{1}{8 R^6 t^3} \left( \sum_{i=1,2,3,4,6} \sum_{n_i} e^{- \frac{\pi n_i^2}{2 R^2 t}}
\sum_m e^{\frac{2 \pi i m Y_5}{R}} e^{- \frac{ \pi m^2}{2 R^2 t}} \right) \right].
\label{z33}
\eea
(\ref{z33}) has an open string ultraviolet quadratic divergence as $t \to 0$. From the closed string picture this is interpreted as the exchange of a massless mode with a 1-pt function in the vacuum. This diagram is reducible in field theory. The D3/$\overline{\hbox{D}}$3 pair both act as sources for massless closed string modes (for example the volume modulus). These modes are unstabilised in the $3\bar{3}$ background and so there is a 1-pt function for them induced by the D3/$\overline{\hbox{D}}$3 pair. As this is a reducible field theory diagram associated to modes present in the massless spectrum we need to subtract off this divergence to extract the brane/interbrane potential.

The interaction potential for a D3/$\overline{\hbox{D}}$3 pair is then given by
\bea
V_{3\bar{3}} & = & \int \frac{dt}{2t} \frac{1}{(2 \pi^2 t)^2} \hbox{Tr}(1) \ti 32 t^4 \ti \left[ \sum_{i=1,2,3,4,6} \sum_{n_i} e^{-2 \pi t n_i^2 R^2}
\sum_m e^{-2 \pi t (Y_5 + m R)^2} - \frac{1}{8 R^6 t^3} \right] \nn \\
& = & \frac{32}{2 (2 \pi)^2} \int dt \, t \, \hbox{Tr}(1) \left[ \frac{1}{8 R^6 t^3} \left( \sum_{i=1,2,3,4,6} \sum_{n_i} e^{- \frac{\pi n_i^2}{2 R^2 t}}
\sum_m e^{\frac{2 \pi i m Y_5}{R}} e^{- \frac{ \pi m^2}{2 R^2 t}} - 1 \right) \right].
\eea
The winding sum effectively contributes between $t \sim R^{-2}$ and $t \sim (2 \pi Y^2)^{-1}$.
When $1 \ll Y \ll R$ we can evaluate the integrals analytically to obtain
\be
\label{33pot}
V = \frac{32}{2(2 \pi^2)^2} N_{D3} N_{\bar{D}3} \left[ \frac{1}{4 \pi^2 Y_5^4} + \mc{O}\left( \frac{1}{8 R^4} \right) + \ldots \right].
\ee
We can normalise this expression by comparing with the standard expression for the D3/$\overline{\hbox{D}}$3 potential (e.g. see \cite{kklmmt})
\bea
\label{33abspot}
V(r) & = & 2 N T_3 \left( 1 - \frac{1}{2 \pi^3} \frac{ N T_3}{M_{10}^8 r^4} \right) \nn \\
& = & 2 N T_3 \left( 1 - \frac{g_s}{4 \pi^3} \frac{N}{(r / 2 \pi \sqrt{\alpha'})^4} \right).
\eea
Here $T_3 = \frac{2 \pi}{g_s (2 \pi \sqrt{\alpha'})^4}$ is the tension of a D3 brane and $\frac{1}{M_{10}^8} =
\frac{g_s^2 (2 \pi \sqrt{\alpha'})^8}{2 \pi}$.

\subsubsection*{D5/$\overline{\hbox{D}}$5 interaction potential}

We now want to modify this calculation to obtain the mutual interaction between D5 branes and $\overline{\hbox{D}}5$ branes that
are separated by a large distance in the Calabi-Yau but wrap homologous cycles, so that there is no overall 5-brane tadpole
on the cycle. We do so by extending the above calculation to orbifold models where we can consider D5 and $\overline{\hbox{D}}5$ branes
wrapping collapsed but homologous cycles. These can be described as fractional D3 branes.

Let us start with a telegraphic review of orbifold singularities and their relationship to fractional brane charges.
The orbifold action can be written as $z_i \to e^{2 \pi i \theta_i} z_i$, and a supersymmetric orbifold requires
$\theta_1 + \theta_2 + \theta_3 = 0$. If all $\theta_i \neq 0$, this is called an `$\mc{N} = 1$' (fully twisted) sector and if $\theta_1 + \theta_2 = 0, \theta_3 = 0$ (or permutations), this is called an $\mc{N}=2$ (partially twisted) sector. The case of $\theta_i = 0$ for all $i$ is called the
`$\mc{N}=4$' (untwisted) sector. A fractional D3 brane at an orbifold singularity in general carries D7, D5 and D3 charge.
This corresponds to the fractional
brane being a boundstate of D7, D5 and D3 branes, wrapping cycles that are collapsed to zero size at the singularity.

At orbifold singularities
the D3, D5 and D7 charges relate closely to $\mc{N}=4$ (untwisted), $\mc{N}=2$ (partially twisted) and $\mc{N}=1$ (fully twisted) sectors of the orbifold.
For $\mc{N}=1$ sectors both the 4-cycle and its dual 2-cycle are collapsed at the singularity, and all tadpole cancellation must take place locally.
This accounts for all the D7 charge plus the D5 charges associated to 2-cycles dual to local 4-cycles.
For $\mc{N}=1$ sectors the 4-cycles and 2-cycles have no distant homological relatives.
For $\mc{N}=2$ sectors, there is a collapsed 2-cycle at the singularity, but the dual 4-cycle is non-compact.
This cycle may be homologous to distant 2-cycles. These correspond to all the D5-charges not accounted for by $\mc{N}=1$ sectors.
$\mc{N}=4$ sectors are associated to bulk tadpoles and correspond to D3 brane charge with no associated cycle.

For our problem of interest we therefore construct a fractional brane consisting of a bound state
that carries both D3 and D5 charge. The D5 charge implies that the fractional brane corresponds to
a D5 brane wrapped on a collapsed 2-cycle. The requirement that the 2-cycle be homologous to a distant 2-cycle, so that
tadpole cancellation need not occur locally, implies
that the D5 charge should lie in the $\mc{N}=2$ sector of the orbifold. We then in addition need to add at a distant
singularity a fractional $\overline{\hbox{D}}3$ brane which is a bound state of $\overline{\hbox{D}}3$ and a $\overline{\hbox{D}}5$ brane wrapping a homologous cycle.

\begin{figure}
\begin{center}
\epsfig{file=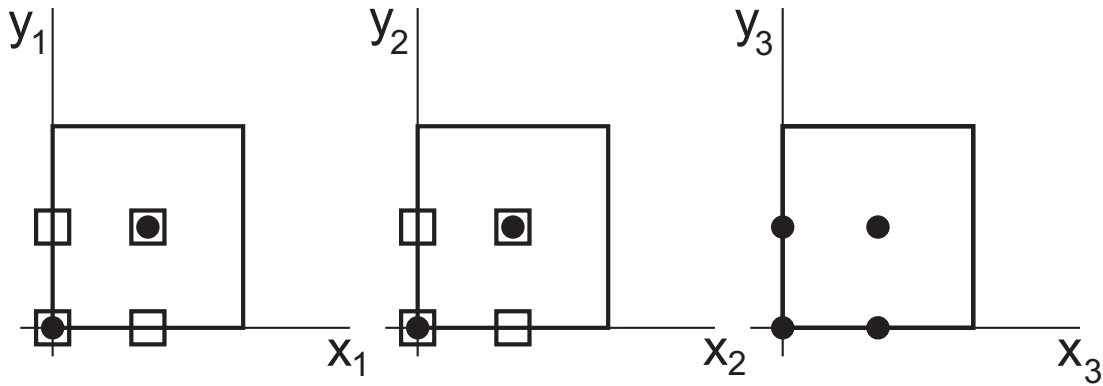,height=5cm}
\caption{The $T^6/\mathbb{Z}_4$ orbifold.}
\label{z4orbifold}\end{center}\end{figure}

This setup can be arranged by working with toroidal orbifolds. Consider for example the $T^6/\mbb{Z}_4$ orbifold, generated by the 
orbifold action $(\theta_1, \theta_2, \theta_3) = (1/4, 1/4, 1/2)$.
The geometry of this space is shown in figure \ref{z4orbifold}.
As a compact space this orbifold has $h^{1,1} = 31, h^{2,1} =7$. The 31 elements of $h^{1,1}$ decompose as 5 untwisted 2-cycles, 16 $\theta^1$ twisted cycles stuck at the 16 $\mbb{Z}_4$ fixed points, 6 $\theta^2$ twisted cycles stuck
at $\mbb{Z}_4$ invariant combinations of $\theta^2$ fixed points, and 4 $\theta^2$ twisted cycles at $\mbb{Z}_4$ fixed points and
propagating across the third $T^2$.

Our main interest is in the last class of four 2-cycles. The point is that for these, for
each $\mbb{Z}_4$ fixed point the $\theta^2$ sector is not in homology
uniquely associated to that fixed point: it is rather shared by the four fixed points differing by their location in the
$(x_3, y_3)$ plane. For example, the fixed points $(0,0,0)$, $(0,0,\frac{R_1}{2})$, $(0,0,\frac{iR_2}{2})$ and $(0,0,\frac{R_1 + i R_2}{2})$
are all connected by a single 2-cycle from the $\mc{N}=2$ sector.
Each $\mbb{Z}_4$ fixed point therefore has a local 2-cycle which is a individual
representative of the cycle homology: however the other $\mbb{Z}_4$ fixed points, distant in the $(x_3, y_3)$ plane,
also have 2-cycles that lie in the same homology class.

We can therefore obtain precisely the desired setup by placing a fractional D3 brane (D3/D5 bound state) at the
$(0,0,0)$ $\mbb{Z}_4$ singularity and a fractional $\overline{\hbox{D}}3$ brane ($\overline{\hbox{D}}3/\overline{\hbox{D}}5$ bound state) at the $(0,0,R/2)$ $\mbb{Z}_4$ singularity.
Such a configuration will cancel all global RR tadpoles and contains a D5 and a $\overline{\hbox{D}}5$ wrapping distant but homologous cycles.
This configuration is also fully amenable to a worldsheet analysis and we can compute the D5/$\overline{\hbox{D}}5$ interaction potential
using precisely the same techniques we used for the annulus computation.

The computation of the partition function on the orbifold essentially just involves the insertion of the orbifold
trace $\frac{(1 + \theta + \theta^2 + \theta^3)}{4}$ acting on the open string spectrum, and also the inclusion of the action
of the orbifold on the Chan-Paton factors.
For a $\mbb{Z}_4$ singularity it is given by
\bea
\mc{A} & = & \int_0^{\infty} \frac{dt}{2t} \, \hbox{STr}\left( \frac{(1 + \theta + \theta^2 + \theta^{3})}{4}
\frac{1 + (-1)^F}{2} \, q^{(p^\mu p_\mu + m^2)} \right).
\eea
Here $q=e^{-2 \pi t}$ and $\hbox{STr} = \sum_{bosons} - \sum_{fermions} \equiv \sum_{NS} - \sum_R$. As before, the partition function
for a brane-antibrane system involves a sign flip for the spins structures corresponding to RR exchange. 

The Chan-Paton matrix is $\gamma_{\theta} = \hbox{diag}(1, \omega, \omega^2, \omega^3)$ with $\omega = e^{2 \pi i/4}$.
We choose a fractional D3 brane configuration of $(N, M, N, M)$ at the $(0,0,0)$ singularity and a fractional $\overline{\hbox{D}}3$
configuration of $(N, M, N, M)$ at the $(0, 0, R/2)$ singularity. This cancels all global RR tadpoles. The D3 charge of the configuration is
$(N + M)/2$, and the D5 charge $(N - M)/2$ (and likewise for $\overline{\hbox{D}}3$ and $\overline{\hbox{D}}5$ charge).

The D3/$\overline{\hbox{D}}3$ contribution to the interbrane potential is determined by the $\mc{N}=4$ sector (1 inserted). The
D5/$\overline{\hbox{D}}5$ contribution is determined by the $\mc{N}=2$ sector ($\theta^2$ inserted). The $N=1$ sectors give no contribution
(they vanish due to tadpole cancellation).

For the D5/$\overline{\hbox{D}}5$ part the relevant partition function is given by
\bea
& & \int \frac{dt}{2t} \frac{1}{(2 \pi^2 t)^2} \hbox{Tr}(\gamma_{\theta^2} \otimes \gamma_{\theta^2}^{-1}) \sum_{\alpha, \beta = 0,1/2}
\eta_{\alpha \beta}
\left( \frac{ \thba{\alpha}{\beta}{it} }{\eta(it)^{3}} \right)^2 \left( \prod_{i=1}^2 (-2 \sin \pi \theta_i)
\frac{\thba{\alpha}{\beta + \theta_i}{it}}{\thba{\half}{\half + \theta_i}{it}} \right) \nn \\
& & \times  \sum_{i=5}^6 \sum_{m_i} e^{-2 \pi t (Y_i + m_i R_i)^2}.
\eea
For the $\mbb{Z}_4$ case, $\theta_i = (1/2,1/2,0)$ in the $\mbb{N}=2$ sector.
Using the modular transformations we find that the small $t$ expansion of the oscillator series is
\be
\hbox{D}5/\bar{\hbox{D}}5 : 32 \sin^2 \pi \theta \left( 1 + \ldots \right),
\ee
where subleading terms are of order $e^{-2 \pi/t}$.

We now perform the same steps for the D5/$\overline{\hbox{D}}5$ case that we did earlier for the D3/$\overline{\hbox{D}}3$ case.
We take $Y_5 = R/2$ and $Y_6 = 0$.
There is again a quadratic divergence as $t \to 0$ which corresponds in closed string channel
to the exchange of the twisted mode that is fixed at the origin for the first two tori and propagates freely on the third torus.
The brane-antibrane background gives this mode a vacuum one-point function.
To extract the interaction potential, we need to subtract off this field theory divergence. Doing so gives
\be
\frac{32}{2 (2 \pi^2)^2} \int \frac{dt}{t} \sin^2(\pi \theta) \hbox{Tr}(\gamma_{\theta^2} \times \gamma_{\theta^2}^{-1})
\left[ \sum_n e^{-2 \pi t n^2 R^2} \sum_m e^{-2 \pi t (Y + m R)^2} - \frac{1}{2 R^2 t} \right]
\ee
This is equivalent to
\be
\frac{32}{2 (2 \pi^2)^2} \int \frac{dt}{t} \sin^2(\pi \theta) \hbox{Tr}(\gamma_{\theta^2} \times \gamma_{\theta^2}^{-1})
\left[ \frac{1}{2 R^2 t} \left( \sum_n e^{-\frac{\pi n^2}{2 R^2 t}} \sum_m e^{2 \pi i m Y/R}  e^{- \frac{\pi m^2}{2 R^2 t}} - 1 \right) \right]
\ee
By analysing the two expressions, we can see that the winding sum in the integrand only contributes for $t \lesssim Y^{-2}$, and that for
$t \lesssim R^{-2}$ the integrand vanishes up to terms exponentially suppressed in $e^{- \frac{\pi}{2 R^2 t}}$.

For the orbifold, the singularities are at fixed location and in principle $Y$ is not a tunable parameter. However to see
the nature of this potential,
let us consider the formal limit of small $Y$, where $1 \ll Y \ll R$.  In this case the integral effectively reduces to
\be
\frac{32}{2(2 \pi^2)^2} \int_{t \sim R^{-2}}^{t \sim (2 \pi Y^2)^{-1}} \frac{dt}{t} \sin^2 \theta \, \hbox{Tr} (\gamma_{\theta^2} \times \gamma_{\theta^2}^{-1}) \left(
e^{-2 \pi t Y^2} - \frac{1}{2 R^2 t} \right).
\ee
Approximating $e^{-2 \pi t Y^2}$ as 1 for $t < (2 \pi Y^2)^{-1}$ and 0 for $t > (2 \pi Y^2)^{-1}$, we obtain
\be
\label{55pot}
V_{5\bar{5}} = \frac{32}{2 (2 \pi^2)^2} \left( \sin^2 \pi \theta \, \hbox{Tr} (\gamma_{\theta^2} \ti \gamma_{\theta^2}^{-1} ) \left[ 2 \ln \left( \frac{R}{Y} \right)
- \ln (2 \pi) + \half + \ldots \right] \right).
\ee
This shows a logarithmic interaction potential that does not vanish at infinity. Recalling that the physics of the situation implies $Y \equiv (R - Y)$, the approximate numerical expression (\ref{55pot}) actually offers a good approximation to the full numerical result across
the entire range of $Y$.

\subsubsection*{Normalisation}

As (\ref{33pot}) and (\ref{55pot}) come from the same string diagram, we can compare them to obtain the
precise relative normalisation of the D3/$\overline{\hbox{D}}$3 and D5/$\overline{\hbox{D}}$5 interaction potentials.
The absolute normalisation comes from comparison with (\ref{33abspot}).
For bound states of branes carrying both D3 and  D5 charge on the collapsed cycle,
we then obtain the leading dependence on $Y$ is
\be
V(r) = 2 T_3 \left( N_{D3} - \frac{g_s}{4 \pi^3} \frac{N_3 N_{\bar{3}}}{(Y / (2 \pi \sqrt{\alpha'}))^4}
- \frac{g_s}{\pi} N_5 N_{\bar{5}} \, \sin^2 \theta \left( 2 \ln \left( \frac{R}{Y} \right) - \ln (2 \pi) + \half + \ldots \right) \right).
\ee
We see that the $5\bar{5}$ interaction potential is dominant over the $3\bar{3}$ interaction potential over essentially all length scales.

\subsection*{Extension to Non-Orbifold Models}

We have derived the logarithmic D5/$\overline{\hbox{D}}5$ interaction potential for the simple case of a $T^6/\mbb{Z}_4$ orbifold. We now explain why this result
will hold more generally.

The physical origin of the logarithmic divergence (and also the $r^{-4}$ behaviour for D3/$\overline{\hbox{D}}$3 models) is simple to understand.
The Green's function for a 2-dimensional plane is $\ln r$, and that for a six dimensional volume $r^{-4}$. The logarithmic potential
arises through the exchange of modes that are restricted to lie on a 2-dimensional subsurface of the 6-dimensional
space.\footnote{A D5/$\overline{\hbox{D}}5$ pair separated in flat space talk only through the exchange of bulk closed string modes, which can propagate on all
four transverse directions, giving the naive $r^{-2}$ D5/$\overline{\hbox{D}}5$ interaction potential.}
Such modes are
easy to identify. They correspond to twisted closed string modes in the $\mc{N}=2$ sector. Using $X^i$ to denote coordinates on a complexified plane, these satisfy for an $\mc{N}=2$ sector
\bea
X^1(\sigma, \tau) & = & e^{2 \pi i \theta} X^1(\sigma + 2 \pi, \tau), \nn \\
X^2(\sigma, \tau) & = & e^{-2 \pi i \theta} X^2(\sigma + 2 \pi, \tau), \nn \\
X^3(\sigma, \tau) & = & X^3(\sigma + 2 \pi, \tau).
\eea
The orbifold identification in the first two tori restricts these modes to a fixed point in the first two complex planes, while the modes
are unrestricted in the third plane. These closed string modes (and their KK copies) effectively propagate only on the third torus.

It is easy to see that this condition holds for any $\mc{N}=2$ sector of an orbifold model: the twisted closed string mode in this
sector can always propagate only on a 2-dimensional subspace of the bulk space, and so back-reacts logarithmically on the space.
This argument holds for any orbifold model, and we can use these to see why this should also hold for more complicated singularities. For example, consider the non-Abelian orbifold $\mbb{C}^3/\Delta_{27}$. As an orbifold, this has eight $\mc{N}=2$ sector conjugacy classes and closed strings
modes sourced in these sectors propagate along 2-dimensional subsurfaces of the bulk space. However $\mbb{C}^3/\Delta_{27}$ is also on the moduli
space of the del Pezzo 8 singularity, and we can identify the eight $\mc{N}=2$ conjugacy classes of the orbifold
with the eight 2-cycles of $dP_8$ whose global homological status is not specified in the local model. The map onto the orbifold then
makes it clear that closed string modes associated to these eight $dP_8$ local 2-cycles
will have effective dimension-2 propagation, and associated logarithmic backreaction, far from the singularity.
As resolution of a singularity will not affect the distant Calabi-Yau geometry, it is also clear that the same logarithmic backreaction
will hold for resolved versions of such singularities.

Another non-orbifold example is that of the Klebanov-Tseytlin solution \cite{0002159}, which is sourced by fractional D3 branes on the conifold
that also carry D5 charge on the collapsed 2-cycle. The conifold 2-cycle satisfies all the conditions of the
orbifold $\mc{N}=2$ sectors. It is a 2-cycle defined in the local geometry, for which the dual 4-cycle is non-compact, and
which can have distant representatives in the same homology class.
Indeed the supergravity metric of the Klebanov-Tseytlin solution runs logarithmically in the UV, consistently with the backreaction
expected from charges in an $\mc{N}=2$ sectors.

The general lesson is that branes wrapping local 2-cycles which have distant 2-cycles in the same homology class give logarithmic
backreaction on the geometry. From a closed string perspective, this arises because the modes associated to these 2-cycles only propagate on
effective 2-dimensional subspaces of the bulk volume.
In contrast to the case of 3-branes, the backreaction then grows with increasing distance rather than decaying as $r^{-4}$.

\section{Application to Axion Monodromy Inflation}

What is the significance of our results for axion monodromy inflation?
Our results above were derived for the case of D5/$\overline{\hbox{D}}5$ pairs (as we needed to use worldsheet conformal field theory).
However the physics of the logarithmic interaction potential depended only 
on the dimensionality of the branes and the type of cycle they wrapped,
and this will carry across to the NS5/$\overline{\hbox{NS}}$5 interaction present in axion monodromy inflation.
We then see that the interaction between the distant but homologous 
NS5 and $\overline{\hbox{NS}}5$ is logarithmic and growing in the separation distance. It therefore
remains large even if the throats are well separated. This logarithmic interaction potential
is analogous to that experienced by D7/$\overline{\hbox{D}}$7 pairs. This is a severe problem for 
ensuring stability of the background metric against backreaction.\footnote{Supersymmetric 7-brane configurations
require the transition to F-theory, an intrinsically strongly coupled theory. For non-supersymmetric D7/$\overline{\bar{D}}7$ models,
little concrete is known.}

In the context of warped throats logarithmic backreaction generates a further problem.
Metric perturbations that are logarithmic in the radius are not normalisable and cause large perturbations
at long distance. The energy density of a brane configuration is in part tied up in the profile of the supergravity fields it sources.
The logarithmic backreaction implies that even if the NS5 is located deep in a warped throat the metric perturbations it produce
will grow towards the UV end of the throat. The source of the backreaction is the charge sourced by the NS5. This charge is not cancelled in the
throat (as it requires the presence of the distant tadpole cancelling brane). As the charge is a topological quantity, it cannot be hidden simply
by warping.\footnote{In a similar way, the charge on an electron is not altered by placing it in an accretion disk deep in a black hole gravitational well.} For example, we could imagine constructing an ADM metric at large distances from the throat. This metric must reflect the charge
carried by the source NS5 brane: and so the perturbations induced by the NS5 brane must remain large at long distances.

Such perturbations will therefore still be large at the point where the throat glues into the bulk, and so
would not be suppressed in the bulk.\footnote{This is not
unrelated to the work of \cite{09123519, 11066165} on supergravity solutions involving $\overline{\hbox{D}}3$ branes.
The $r^{-4}$ falloff of the 3-brane metric makes it reasonable to hope that the effects of a $\overline{\hbox{D}}3$ can be localised down a throat
without sourcing non-normalisable perturbations in the UV. In contrast, with a 5 brane we are dealing with $\ln r$ \emph{growth} in the
long distance metric perturbations with no long-distance falloff.}
This suggests that the energy scale of the NS5/$\overline{\hbox{NS}}$5 interaction potential will be set by the UV or bulk end of
the throat, rather than the IR end. In a model (such as suggested in \cite{09072916}) where both throats lie within a single warped region, this similarly suggests that the energy scale of the NS5/$\overline{\hbox{NS}}5$ pair should be set by the highest UV scale needed to connect the two throats. As the inflationary model relies on the presence of warping to suppress the 5-brane/anti-5-brane energy scale to that lying at 
the IR end of the throat, in either case such an effect would destabilise the model.

There is a more directly stringy way to argue this point. As we have discussed, the completely honest string theory way to compute a 
brane-antibrane interaction potential is via the annulus amplitude, which is schematically
\be
\label{kkk}
\mc{A} \sim \int \frac{dt}{t} \hbox{Tr} \,\, e^{-m^2 t},
\ee
where the trace is over all states of the string. The D5 and $\overline{\hbox{D}}$5 branes are geographically separated
by a distance $R (2 \pi \sqrt{\alpha'})$, where $R$ is the dimensionless bulk radius and $\sqrt{\alpha'}$ the bulk string scale.
All open strings contributing to this amplitude must stretch between the D5 and $\overline{\hbox{D}}$5 and therefore have a mass scale set by 
the bulk string scale, $R/\sqrt{\alpha'}$.  

For a full Calabi-Yau compactification with warped throats the full string spectrum cannot be computed. However the above simple geometric argument tells us that the region of moduli space where the potential is generated is $t \sim (R / \sqrt{\alpha'})^{-2}$. This corresponds to string modes with energies set by the bulk scale and which are dominantly located outside the throats. The precise evaluation of (\ref{kkk}) is model
dependent. However the only scale in the problem is the bulk scale and so the size of the brane-antibrane interaction potential should be set by the bulk (or more generally, the mass scale of strings that connect the brane and antibrane) and not warped down by the scale of the throats.

Note that this argument can also be applied to the case of a $\overline{\hbox{D}}$3 brane. However in this case the throat itself carries D3-brane charge via the flux. There is not a direct equivalent of the annulus diagram (open strings do not end on flux). However morally speaking we can view the fluxed throat as a stack of $N$ D3-branes. In this case the scale of the stretched `flux-antibrane' open string that enters the 
annulus diagram is set by the infrared scale of the throat, as the antibrane charge is cancelled at the tip of the throat. This then 
implies that the scale of the $\overline{\hbox{D}}$3 potential is set by the infrared throat scale, in agreement with the analysis of 
\cite{Dymarsky}.

From this viewpoint, the key difference between anti-5-branes and anti-3-branes is that tadpole cancellation for the anti-5-brane requires us to go 
outside the throat. The mass scale of the stretched open strings that enter the annulus diagram is then set by the bulk string scale. 
In contrast, the anti-3-brane charge is cancelled within the throat, and the mass scale of the stretched open strings is set by the IR scale of the throat.

\section*{Conclusions}

This note has argued that 5-brane/antibrane pairs wrapping distant but homologous cycles have a logarithmic interaction potential.
This holds for all cases accessible to a CFT analysis.
The backreaction of such objects is in effect codimension 2, and as for D7 branes 
grows logarithmically with the distance from the brane, remaining
large at long distances. This very large backreaction makes it difficult to localise the effects and energy densities 
of such 5-branes down a warped throat. This represents, at the very least, a model building challenge for models such as axion monodromy inflation which rely on being able to do this.

\acknowledgments{I thank the organisers and participants at the pre-Strings workshop in Stockholm in June 2011 for a stimulating
meeting where these thoughts germinated. I am also very grateful to Cliff Burgess, Liam McAllister and Enrico Pajer
for reading and commenting on the manuscript. I am funded by the Royal Society and Balliol College.}

\end{document}